\documentclass[usenatbib]{article}   
\usepackage{times,graphicx,natbib,a4wide}

\begin{document}

\title{Slingshot Dynamics for Self Replicating Probes and the Effect on Exploration Timescales}
\author{Arwen Nicholson$^1$, Duncan Forgan$^1$}
\maketitle

\noindent $^1$Scottish Universities Physics Alliance (SUPA), Institute for Astronomy, University of Edinburgh, Blackford Hill, Edinburgh, EH9 3HJ, UK \\

\noindent \textbf{Word Count:5,400} \\

\noindent \textbf{Direct Correspondence to: Arwen Nicholson arwen.e.nicholson@gmail.com} \\

\newpage

\begin{abstract}

Interstellar probes can carry out slingshot manoeuvres around the stars they visit, gaining a boost in velocity by extracting energy from the star's motion around the Galactic Centre.  These maneouvres carry little to no extra energy cost, and in previous work it has been shown that a single Voyager-like probe exploring the galaxy does so 100 times faster when carrying out these slingshots than when navigating purely by powered flight  (\citealt{probe_staticbox}).  We expand on these results by repeating the experiment with self-replicating probes.  The probes explore a box of stars representative of the local Solar neighbourhood, to investigate how self-replication affects exploration timescales when compared with a single non-replicating probe.

We explore three different scenarios of probe behaviour: i) standard powered flight to the nearest unvisited star (no slingshot techniques used), ii) flight to the nearest unvisited star using slingshot techniques, and iii) flight to the next unvisited star that will give the maximum velocity boost under a slingshot trajectory.

In all three scenarios we find that as expected, using self-replicating probes greatly reduces the exploration time, by up to three orders of magnitude for scenario i) and iii) and two orders of magnitude for ii). The second case (i.e. nearest-star slingshots) remains the most time effective way to explore a population of stars.  As the decision-making algorithms for the fleet are simple, unanticipated ``race conditions'' amongst probes are set up, causing the exploration time of the final stars to become much longer than necessary. From the scaling of the probes' performance with star number, we conclude that a fleet of self-replicating probes can indeed explore the Galaxy in a sufficiently short time to warrant the existence of the Fermi Paradox.

\end{abstract}

\section{Introduction}

The Fermi Paradox is one of the oldest and most important problems in the Search for Extraterrestrial Intelligence (SETI). The Paradox is defined as the absence of extraterrestrial intelligence (ETI) in our galaxy despite calculations suggesting that galactic colonisation should be feasible within the age of the galaxy (see reviews by \citealt{BrinG.D.1983, Webb2002,Cirkovic2009}). 

The characteristic time for colonisation of the galaxy (either in person or through conventional or self-replicating probes) is known as the Fermi-Hart timescale, calculated by \citet{Hart1975} to be:

\begin{equation}
\mathit{t}_{\mathrm{FH}} = 10^{6} - 10^{8} \, \mathrm{years}.
\end{equation}

As the accepted age of the Earth as an object of roughly present-day mass is $t_{\oplus} \sim 10^9$ years, the large discrepancy between these timescales leads us to the Fermi Paradox.

The first serious proposal to use probes to explore the galaxy is credited to \citet{Bracewell1960} as an alternative to interstellar radio communication between civilisations spread through the galaxy, surmounting the potentially high energy cost of transmitting an easily receivable signal over very large distances \citep{Benford2010,Benford2010a}, and the problems associated with synchronicity of the civilisation lifetimes \citep{SKA,Horvat2011}.  These probes would lurk in potentially habitable planetary systems, to wait for signs of intelligent life.

It is argued by \citet{Freitas1983} that probes may have visited our solar system and that we have been too optimistic regarding our ability to detect the probes \citep{Freitas1983a}.  Freitas argues that objections to the existence of extraterrestrial intelligence based on the Fermi Paradox are invalid, as they are based on unsupported assumptions that ETI or their artifacts are not currently present in our Solar System. He argues that humanity's ignorance to potential evidence of ETI is not appreciated (but see \citealt{Loeb2012}), and notes that any objects sent by aliens not intended to be found, will not be found. He proposes a scenario where a probe camouflages itself so as to set up a threshold test of the technology or intelligence of the recipient species, where the test must be met before the species is allowed to communicate with the device.  Evidence in the form of `spent' or destroyed probes is less likely, as any civilisation attempting interstellar exploration are presumably skilled engineers, and would send probes with the ability to self-repair due to the large travel distances and times required for such a task, giving the probes a very long life-span \citep{Freitas1983}.

Recently, computational power has increased sufficiently to allow the creation of detailed simulations of probes exploring a sector of the Milky Way. \citet{Bjork2007} investigated the timescale for a given number of space probes to explore 40,000 stars in a box, and then modelled 260,000 of these 40,000 stellar systems all located in the Galactic Habitable Zone (GHZ) \citep{GHZ}.  In these simulations the speed of the probes was assumed to be 0.1\textit{c}. This speed was selected as it is low enough to be able to ignore effects of general relativity, but high enough that the time travel is of the order of years for travel between stars. Their results found that with 8 probes with 8 subprobes each, ~ 4$\%$ of the galaxy could be explored in 2.92 x 10$^{8}$ years. With 200 probes again with 8 subprobes each the exploration time was reduced to 1.52 x 10$^{7}$ years. To put these numbers into perspective: it took 2$\%$ of the age of the universe for 8 probes with 8 subprobes to explore $\approx$ 3.85\% of the galaxy so exploring the entire galaxy with this method would take a huge amount of time.

\citet{Cotta2009} extended this analysis by attempting to optimise the trajectory of the probes, an instance of the NP-hard ``travelling salesman problem'' \citep{Golden1988, Toth2001}  in an attempt to determine the possible number of extraterrestrial technological civilisations (ETCs) that could be exploring the galaxy while still giving a high probability that we would not have been visited yet. They found that it was unlikely for more than 10$^{2}$ - 10$^{3}$ ETCs would be exploring the galaxy in a given Myr.  This result assumes a constant probe velocity of $0.1c$, and a probe lifetime of 50 Myr, with contact evidence lasting 1 Myr. The number of ETCs drops considerably if the contact evidence is assumed to last for 100 Myrs. In this case the upper-bound for the number of exploring ETCs goes down to about 10. The method for exploration was again using a number of probes with subprobes. They found that the exploration time was again very high; around 12 Myrs for 4 probes and 3 Myrs for 8 probes to explore a quarter of the galaxy. They conclude by saying that only with a large number of ETCs exploring the galaxy would it become improbable for us to have no evidence of such an exploration. \cite{Cartin2013}'s modelling  shows that even in attempts to explore the Solar neighbourhood, with a fixed number of probes launched as a fleet from the Solar System, can be frustrated by failure of probe components.   Further, these issues cannot be completely avoided by simply increasing the number of probes in the initial fleet.

This implies that for the Fermi Paradox to truly hold, a very large fleet of probes would be necessary - the most efficient means of generating such a fleet invokes the concept of a self-replicating probe (SRP).  So far, no simulations of this type have been carried out using SRPs. Also known as Von Neumann probes, SRPs are spaceships capable of producing an exact copy of themselves under full automation.  If probes require a great deal of self-repair, then it is not unlikely they may be able to achieve the next step of self-replication.   A civilisation may send out one to a few SRPs, which will be programmed to choose the next star they travel to according to some decision making algorithm. Once they reach the new star system, they scan for signs of life, and create a copy of themselves.  The parent and child probe each pick a new star (not the same star) to travel to, and the process repeats itself.

\citet{Freitas1980} outlines the specifics of a SRP, estimating the mass of each component of the spacecraft, and the material the spacecraft would need to accumulate to be able to produce a copy of itself and provide fuel. They suggest that a spacecraft may capture comets and use atmospheric mining to acquire these materials. They state that in their minds there is ``little doubt that such a machine can, in theory, be designed''.

\citet{Wiley2011} presents many of the arguments given against the use of SRPs, concluding that the arguments are insufficient for SRPs to be excluded from attempts to answer the Fermi Paradox.   \citet{Sagan1983} proposed that ETI would be unlikely to use SRPs due to the risk of mutations in the replication process resulting in dangerous unforeseen consequences. For example, a mutated probe may no longer possess the same goals as a ``healthy'' probe, malfunctioning in a process analagous to cancer in biological systems, filling the galaxy with malignant probes and potentially destroying that which it was designed to find.

A counter argument to this can be found in \citet{Tipler1980}.  Biological evolution is a process with no design or foresight, whereas self-replicating machines will be the result of careful intelligent design.  Evolution has developed many measures to protect against malignant mutations just as human technology has built in safeguards against corruption in electronic systems, so it is reasonable to assume that SRPs would be protected in a similar, and possibly better fashion.  While we feel that \citet{Wiley2011}'s argument against dangerous mutation is overly stated in places (for example, he compares the number of probe replications to the number of cell divisions in the human body, which we feel is an incorrect comparison, as the probes are likely to have a large number of cell-analogue components, as well as complex firmware/software), we agree that the concerns of mutations are insufficient to exclude all civilisations from building SRPs.  

A related argument against SRPs includes the  ``predator and prey'' scenario \citep{Chyba2005} where mutated probes abandon their original mission of exploring the galaxy and start to prey on normal probes, decreasing the number of explorer probes and greatly increasing the exploration time. Wiley's counter argument makes the obvious point that although biology gives us many examples of how predators greatly influence the numbers of prey, the SRPs imagined do not behave as normal prey. The probes are continuously travelling at maximum speed and constantly dispersing radially. This means that as the predator probes would have the same maximum velocity as non mutated probes, they would have a hard time catching their prey. It seems implausible to ignore SRPs based on this argument, although \citet{Wiley2011} notes that once a section of the Galaxy is explored, the lurking explorer probes leave themselves open for consumption by the predators, which may prevent some target species from being contacted.

Recent advances in 3D printing suggest self-replicating machines of sufficient complexity may be within our grasp in the coming centuries.  With that in mind, when considering extra-terrestrial civilisations potentially a thousand times older than ours (especially given that terrestrial planets are likely to be 1 Gyr older than the Earth, \citealt{Line_planets}), it is not unlikely that they would have such technology. 

In this paper, we present results from the first models to simulate the colonisation of a sector of the Milky Way using SRPs with realistic probe dynamics.  We build on previous Monte Carlo Realisation (MCR) simulations by \citet{probe_staticbox}, which considered a single non-replicating probe using orbital slingshot maneouvres to provide velocity boosts instead of relying purely on powered flight, as has previously been the case.  Using slingshots inside gravitational potential wells allows a probe to produce relatively large $\Delta v$ and alter its trajectory without expending fuel, boosting its speed relative to the rest frame of its starting position \citep{Surdin1986, Gurzadyan1996}.

In section \ref{sec:method} we describe the construction of the numerical model, in section \ref{sec:results} we show the results of the simulations, in section \ref{sec:Discussion} we note some issues and limitations of the model, and in section \ref{sec:conclusions} we summarise the work.

\section{Method \label{sec:method}}

The simulation domain is composed of stars uniformly distributed with a uniform density of 1 star per cubic parsec. To simulate the rotation curve of the Milky Way the stars were set in a shearing box configuration, however for convenience the stars remain fixed in position even though they have velocity vectors. We created 10 Monte Carlo Realisations for three scenarios, investigating four different values for the total number of stars (see below), giving a total of 120 realisations.  We average each set of 10 realisations to characterise the effect of random fluctuations.  The maximum velocity of the probes was chosen to be $3 \times 10^{-5}c$, where $c$ is the speed of light in a vacuum) which is approximately the maximum velocity obtained by unmanned terrestrial spacecraft such as Voyager 1\footnote{http://voyager.jpl.nasa.gov/mission/weekly-reports/index.htm}.

\subsection{Choosing the Next Star}

We explore realisations of the same three scenarios as \citet{probe_staticbox}:

\begin{enumerate}
\item Powered flight to the nearest neighbour. Each probe will travel to its closest neighbour at its maximum powered velocity. $\Delta v$ is fixed by the repeated deceleration and acceleration the probe makes at every star it visits (labelled \textbf{powered}).
\item Slingshot assisted flight to the nearest neighbour. The path is identical to that for scenario 1, but the probe only accelerates to maximum velocity once and does not decelerate. It instead uses slingshot manoeuvres to repeatedly boost its maximum velocity. It is assumed that the $\Delta v$ required of the probe to make course corrections in order to use a slingshot trajectory is negligible (labelled \textbf{slingshot}).
\item Slingshot assisted probe selecting the next star by seeking the maximum velocity boost. This time the probe seeks the path such that the relative velocity between the current and destination stars is large and negative, meaning the destination star is moving towards the current star. This will result in a larger velocity boost, but in general will require a longer path length to achieve it (labelled \textbf{maxspeed}).
\end{enumerate}

We apply these scenarios to self-replicating probes to explore how self-replication affects exploration time for each scenario.

\subsection{The Dynamics of Slingshot Trajectories}

The slingshot trajectories used in these simulations are the same as those described in \citet{probe_staticbox}. A slingshot trajectory uses the momentum of the star it passes to either lose or gain velocity depending on the incident angle of the probe's approach. This means the probe does not need to use the additional energy that would be required to perform a similar trajectory using only powered flight, and may receive a velocity boost. Here we briefly cover the mathematics of slingshot trajectories. For more detail the reader is referred to \citet{Gurzadyan1996}, in particular Chapter XIII, section 4.

\begin{figure}[htbp]
\begin{center}
\includegraphics[scale=0.5]{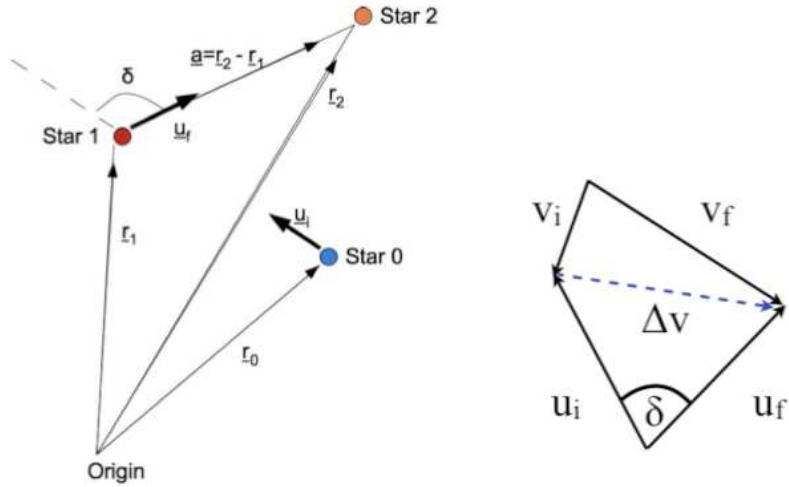}
\caption{The slingshot maneouvre. The probe changes its direction by an angle $\delta$ while the magnitude of its velocity, in the frame of star 1, $\mathbf{u}$, remains constant. Diagram taken from \citet{probe_staticbox} .}
\label{fig:slingshot_trajectory}
\end{center}
\end{figure}

The left side of Figure \ref{fig:slingshot_trajectory} shows one stage of a probe's trajectory using slingshots. From its starting point at Star 0, the probe accelerates under its own power to velocity $\textbf{u}_{i}$ (measured in the reference frame where Star 1 is at rest).  Once it has arrived at star 1, the probe achieves a velocity change $\Delta \mathbf{u}$ during the slingshot such that it leaves Star 1 with velocity $\textbf{u}_{f}$ (again in Star 1's reference frame). $\delta$ is the angle between $\textbf{u}_{i}$ and $\textbf{u}_{f}$. The probe follows a hyperbolic trajectory while performing this manoeuvre and the speed of the probe in this frame remains unchanged:

\begin{equation}
|\mathbf{u}_{i}| = |\mathbf{u}_{f}| 
\end{equation}

Due to the huge mass difference between the probe and the star, the transfer of momentum from the star to the probe is negligible. In the simulation, we observe the probe in the frame where the Galactic Centre is at rest.  The initial and final velocities in this frame are $\mathbf{v}_i$ and $\mathbf{v}_f$ respectively (right side of Figure \ref{fig:slingshot_trajectory}).  The magnitude of the two vectors in the Galactic frame are no longer equal, and

\begin{equation}
\Delta v = 2|\mathbf{u}_{i} | \mathrm{sin}   \left(  \frac {\delta}{2}  \right)
\end{equation}

The probe's initial velocity between Star 0 and Star 1 is the maximum velocity it can achieve under powered flight, but as it travels to more stars using slingshot manoeuvres it can increase its speed. The magnitude of the boost attained by a slingshot manoeuvre is greater if the star's velocity is parallel to that of the probe's trajectory. Thus it is possible for a probe to choose a course based on the proper motion of stars relative to one another such that it performs slingshots with maximal $\Delta v$.

\subsection{Probe Replication}

Our model for probe replication is a very simple one. While a probe is travelling between stars, we assume it collects matter from the interstellar medium, and this is used to create a replica probe. We assume that the quantity of material collected during the flight is great enough such that the probe does not have to stop and mine for materials at any time, and thus the process of replication does not affect the journey time between any two stars.

Each new probe is an exact copy of its parent probe, and no mistakes in the building or programming of new probes are ever made. Thus, each new probe behaves in exactly the same way as the original probe; they select the next star to visit in the same way as their parent, and use (or don't use) slingshot trajectories in the same way.

We assume that a probe releases its replica on arrival at the destination star and the probe does not slow down or change its motion in any way while releasing the replica probe. This allows us to use the same equations as in the single probe case, as the release of a new probe makes no change to the parent probe's motion in any way. We also assume that this replica probe is released with the same velocity and momentum as its parent probe. 

This means, in the powered case, the parent probe slows as always on approach of the new star and then releases the replica with the same velocity. Both probes then choose a new destination star and accelerate away towards their chosen destinations.

In the slingshot case, the parent probe reaches the new destination star, and before it slingshots around the star it releases the replica probe. Both the parent and replica use the slingshot to boost their velocity. As the velocity boost from a slingshot trajectory depends on the angle between the stars, the parent and the replica will achieve different velocity boosts as they will have different destination stars.

\subsection{Information Transfer}

In a situation with more than one probe exploring a galaxy there needs to be a way for the probes to communicate which stars have already been visited. If the probes do not communicate then we can easily end up in a situation where two or more probes land on the same star, and as they all have the exact same algorithm for selecting their next destination star, they may end up following the same path for the rest of the exploration. This could result in a fleet of probes following the same path around the galaxy, which will not only increase the total exploration time, but could also cause chaos if said fleet ended up in a star system with life. One probe visiting may be of enormous consequence to a civilisation \citep{Shostak2002,Almar2011}, but if thousands of probes all descend at once or within a short time of each other it could be interpreted as hostile, causing panic (cf \citealt{Cantril1940}).

We prevent this from happening in our simulations by giving each probe perfect information at all points in time. That is, all probes know exactly where every other probe has been and is going to. When a probe selects its next destination star, it will select an unvisited star that no other probe is currently travelling to. Once a probe selects a destination star, all other probes know not to visit that star.

This situation is obviously unrealistic. While the model might be accurate at the start of the journey when the probes are relatively close together, as they get further and further apart, the information transfer between them should take longer. The rate of transfer of information is limited to the speed of light, so with greater distances there is a greater time delay between a probe making a decision, and the rest of the probes being made aware of this decision. However, we do not take account of this problem and just use the simple case of all probes knowing everyone's actions at all times.  In later sections, we will describe some problems that arise using this assumption.

\section{Results \label{sec:results}}

\subsection{Comparing Single Probes to Self-Replicating Probes}

\begin{figure*}
\begin{center}$
\begin{array}{c}
\includegraphics[scale=0.5,clip=true, trim= 1 1 1 1]{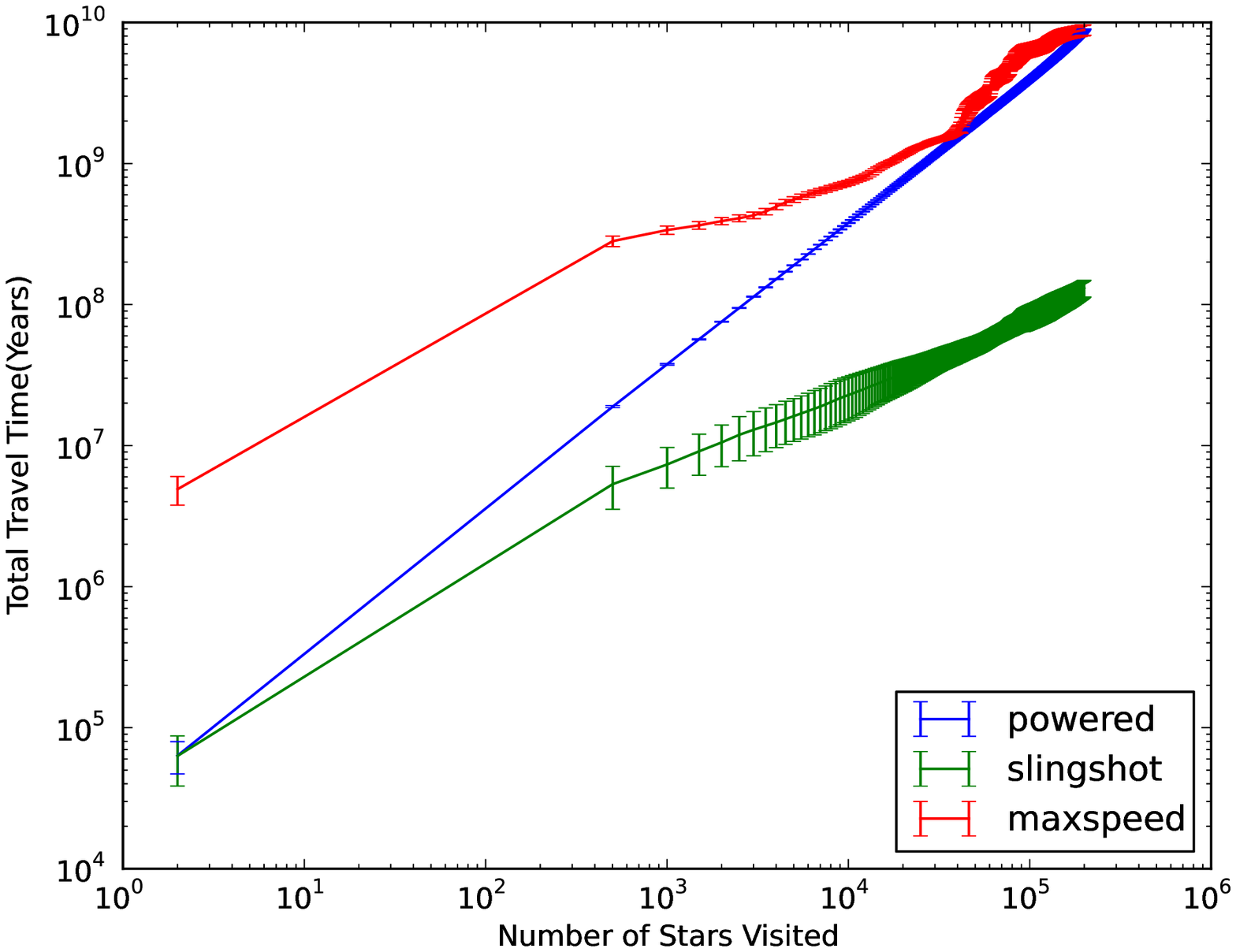} \\
\includegraphics[scale=0.5, clip=true, trim= 1 1 1 1]{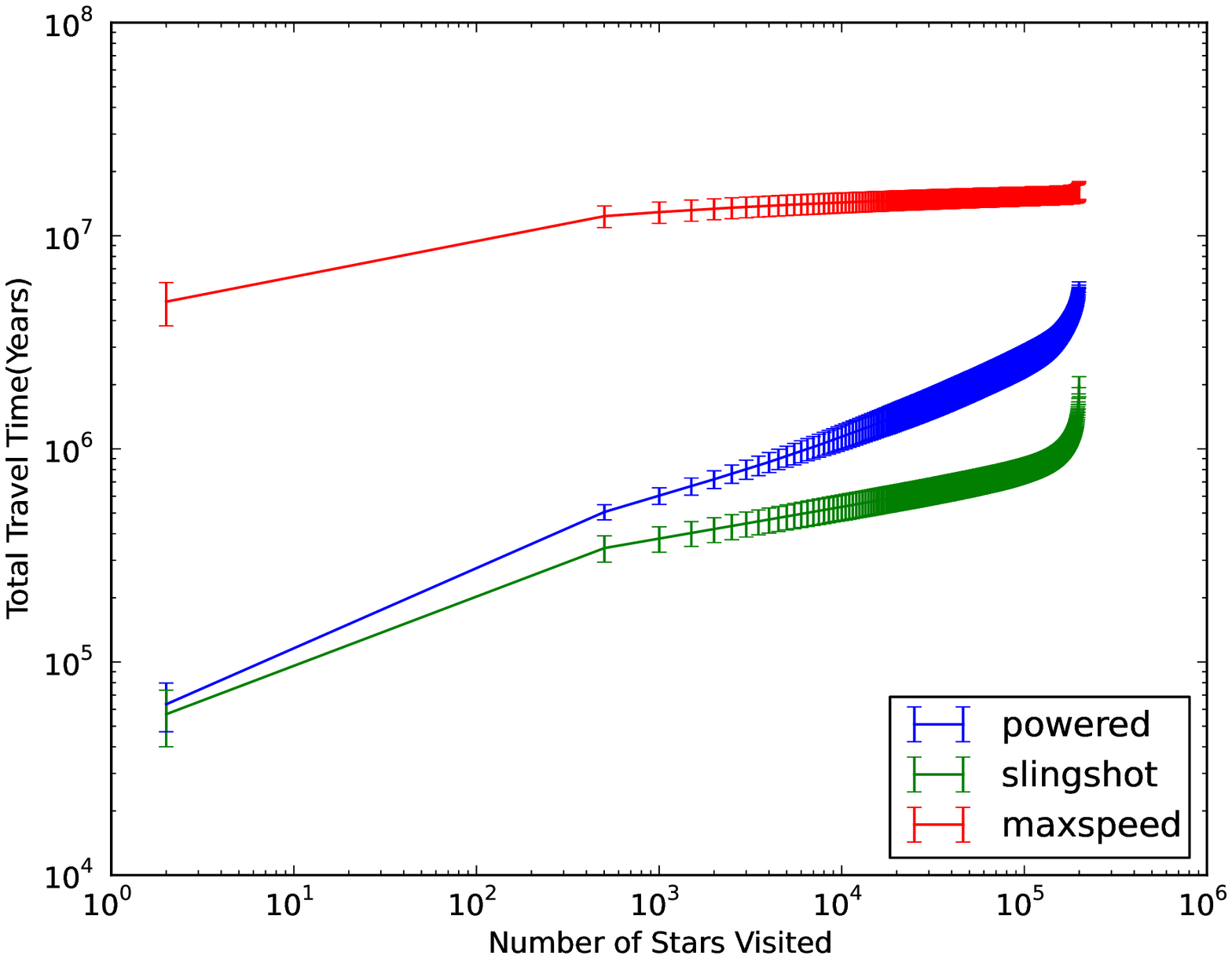} \\
\end{array}$
\caption{The mean cumulative travel time for the 3 scenarios powered, slingshot and maxspeed as described previously, in a box containing 200,000 stars.  The top plot shows the travel times for a single, non-replicating probe, and the bottom plot shows the travel time in the case of self-replicating probes. Every 500th data point was plotted to prevent overcrowding of the graph. The thickness of the lines indicate sample standard deviations as error estimates.  The maxspeed case is represented in red, powered in blue and slingshot in green.}
\label{fig:single_replicate_compare}
\end{center}
\end{figure*}

In general, using self-replicating probes significantly reduces exploration time for all three scenarios when compared with a single non replicating probe.  The top plot in Figure \ref{fig:single_replicate_compare} shows the average cumulative travel time for single non-replicating probes exploring a box of 200,000 stars in the three scenarios, and the bottom plot in Figure \ref{fig:single_replicate_compare} shows the same quantity for the self-replicating probes.  It is immediately clear that adding self-replication - simply by virtue of dividing the task of exploration between many agents - significantly reduces the exploration timescale.  For the propermotion and powered scenarios the exploration time is reduced by a factor of 1000, from $\sim 10^{10}$ years to $\sim 5\times 10^{6}-10^{7}$ years, while the exploration time for the slingshot scenario is reduced by a factor of around 100, from $10^{8}$ years to $10^{6}$ years.  Note that the ordering of the three scenarios remains the same - slingshot is the most effective algorithm, with powered now a close second to the maxspeed scenario.  As the slingshot scenario is already quite efficient, the factor by which adding self-replication reduces exploration timescales is quite low, but in general the effect of self-replication is much more important than the effect of probe dynamics.  We discuss the dependence of this efficiency increase on star number in the following section.

Note that the slingshot and powered scenarios in the self-replicating case show a significant upturn in travel time as the number of stars explored approaches 100\%. A similar, but weaker feature is present in the single probe case for powered flight (see Figure 2 of \citealt{probe_staticbox}), which occurs when there are few unvisited stars left in the domain, with generally larger spatial separation between them than the mean stellar separation.  We might naively think that this problem should no longer exist for the case of self-replicating probes.  In this late stage of exploration, with many hundreds of thousands of probes, each unvisited star should be surrounded by a large number of potential visitors.  The fact that this phenomenon persists is indicative of a deeper problem of modelling inter-probe communication (see Discussion).

\begin{figure}
\begin{center}
\includegraphics[scale=0.46]{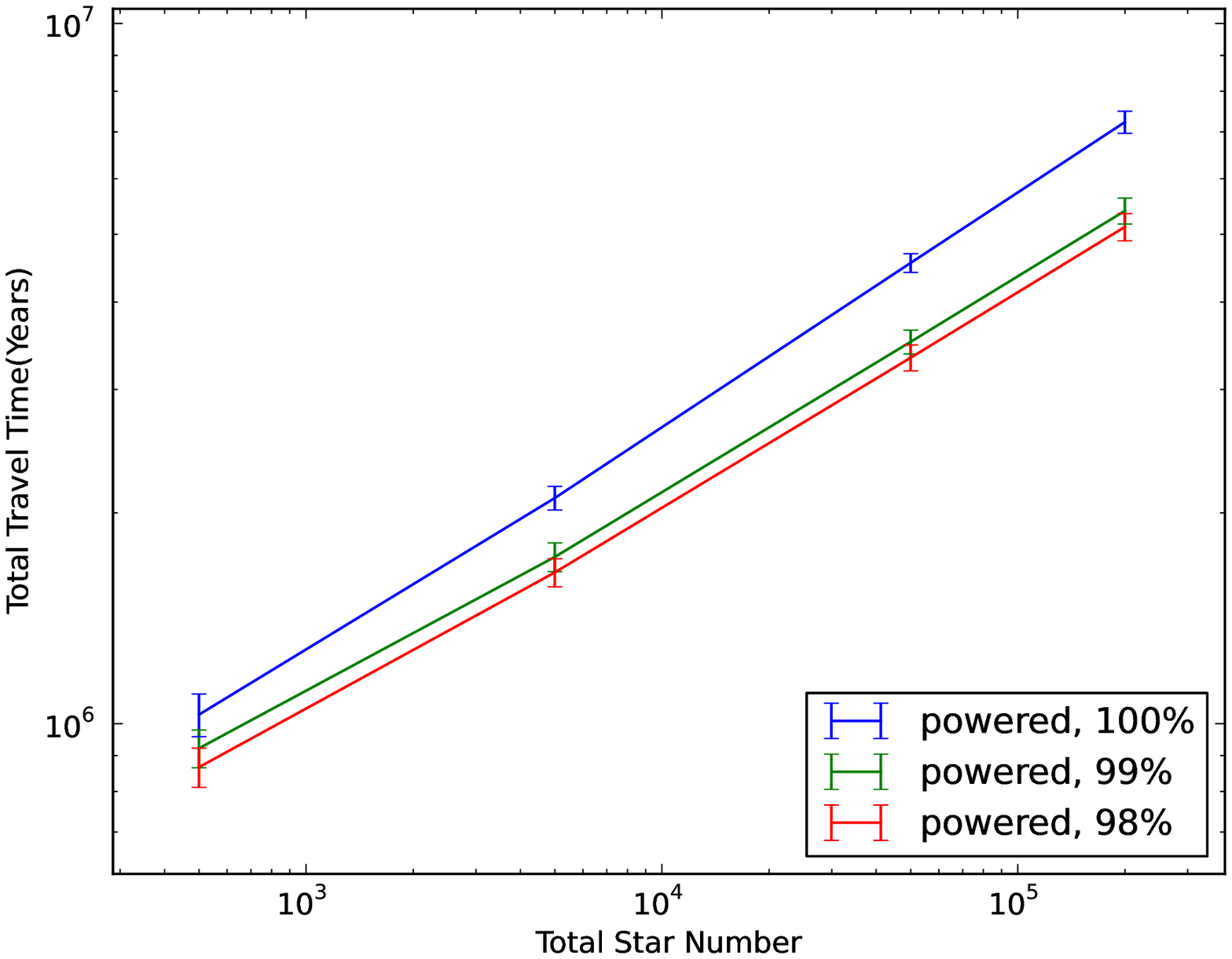}
\caption{Comparison of the exploration time for self-replicating probes to explore 100\%, 99\% and 98\% of the stars in the powered case.}
\label{fig:powered_percent}
\end{center}
\end{figure}

\begin{figure}
\begin{center}
\includegraphics[scale=0.46]{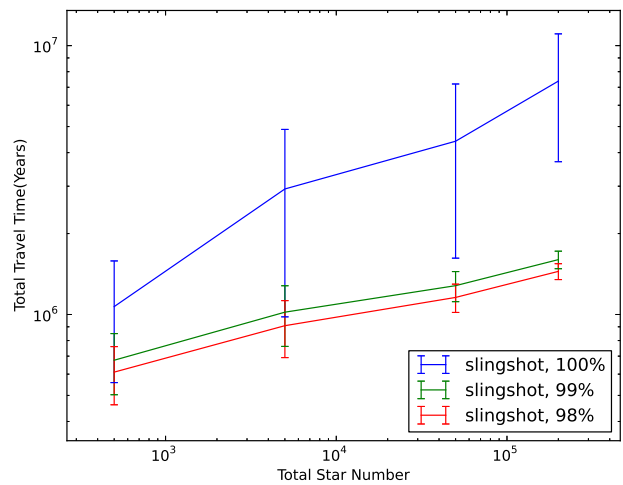}
\caption{Comparison of the exploration time for self-replicating probes to explore 100\%, 99\% and 98\% of the stars in the slingshot case.}
\label{fig:slingshot_percent}
\end{center}
\end{figure}

\begin{figure}
\begin{center}
\includegraphics[scale=0.46]{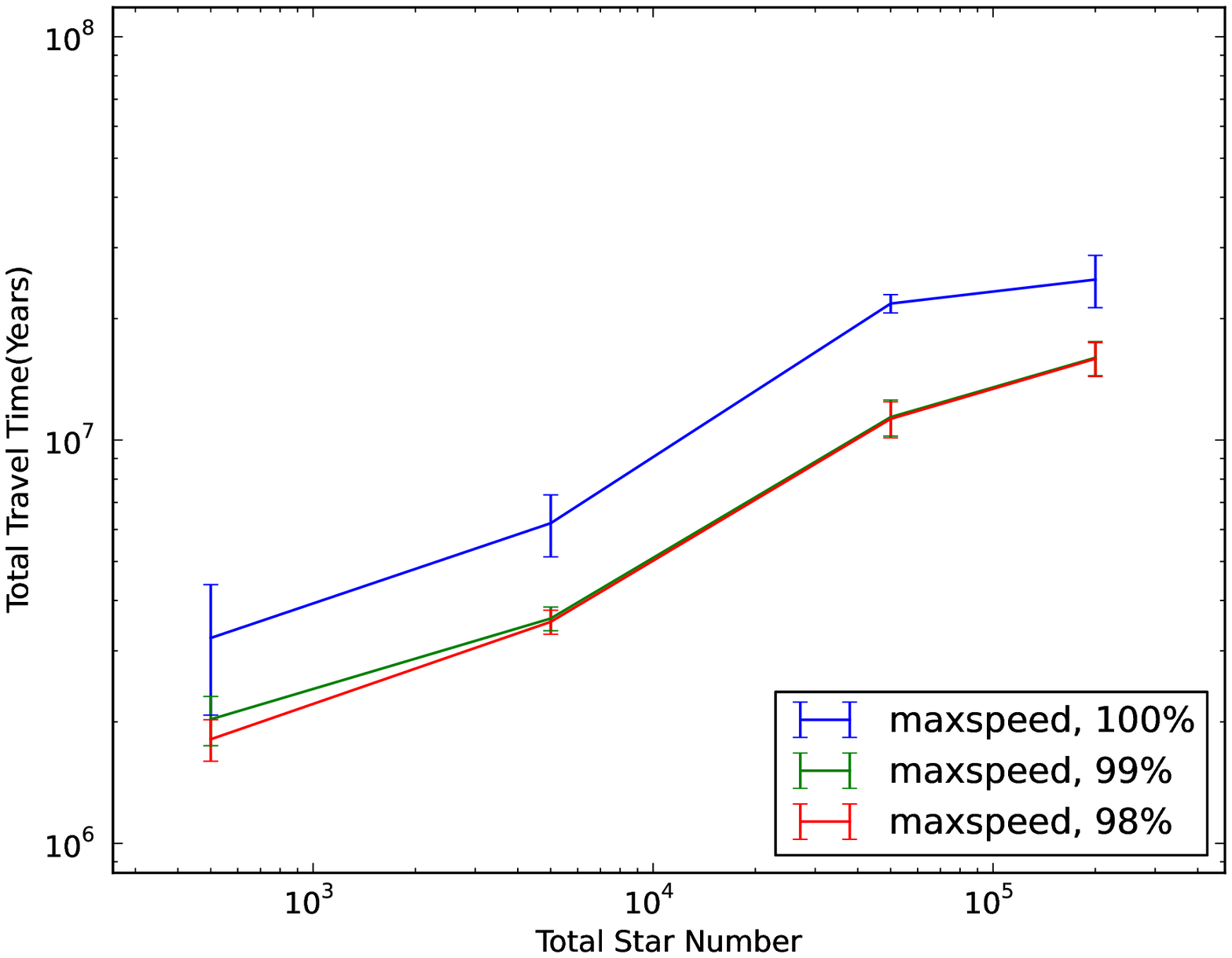}
\caption{Comparison of the exploration time for self-replicating probes to explore 100\%, 99\% and 98\% of the stars in the maxspeed case.}
\label{fig:maxspeed_percent}
\end{center}
\end{figure}

We can assess this phenomenon by investigating the time taken to explore 98\% and 99\% of the domain for the three scenarios as a function of total star number (Figures \ref{fig:powered_percent}, \ref{fig:slingshot_percent}, \ref{fig:maxspeed_percent}).  In the powered case (Figure \ref{fig:powered_percent}), there is a clear power-law relationship between the number of stars in the domain and the total travel time.  The time taken to explore 98\% of the system is the same as the time taken to explore 99\% of the system (to within the sample standard deviation).  The time taken to explore the final 1\% of the domain increases the final travel time by around 40\%!

In the slingshot case (Figure \ref{fig:slingshot_percent}), we can see that the travel time increases more slowly with extra star number - the velocity boosts due to the slingshots allow the probes to explore at greater speed than in the powered case.  Again, the 98\% and 99\% curves are tightly correlated with overlapping errors, but the 100\% case takes nearly three times longer to explore the domain for 200,000 stars, with much larger errors than the powered case due to the larger range of speeds that the probe population can adopt.

The maxspeed case (Figure \ref{fig:maxspeed_percent}) shows more complicated behaviour.  In low star number runs, the time penalty for travelling large distances to achieve a high velocity boost is weaker, as the box itself is smaller.  As the box size increases, the penalty for finding the maximum boost increases, and begins to level off once the star number reaches 100,000, where the time penalty is approximately balanced by achieving a high velocity quickly.  Once more, we can see that the last 1\% of the stellar population requires half the total travel time to explore.

\section{Discussion \label{sec:Discussion}}

\subsection{Race Conditions, Information, and The Final 1\% of the Stars}

\begin{figure}
\begin{center}
\includegraphics[scale=0.5]{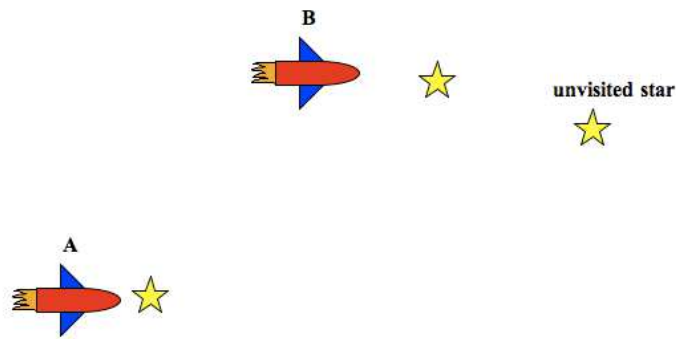}
\caption{An example of a race condition in the simulation.  Although probe B is much closer to the unvisited star, as probe A will reach its destination first, it will tell all other probes it is going on to the unvisited star, thus forcing B to chose another more distant star (or causing B to stop if we are nearing the end of the exploration) which can have the effect of increasing the overall exploration time.}
\label{race_condition}
\end{center}
\end{figure}

\noindent In assuming the probes are perfectly informed of the swarm's exploration of the box, we unwittingly created a problem for the final 1\% of the unexplored box. When probes select their next target, this is done on a first-come first-served basis. The first probe to lay claim to a star will visit it, and all other probes will know immediately not to also choose that star. When there are plenty of stars to choose from, this does not cause any visible issues.  This is only a problem if there are few stars to choose from, where we can end up with the scenario as depicted in Figure \ref{race_condition}. In this simplified situation there are three stars and two probes. Each probe is assigned a star: the probe to reach their star first will go on to visit the final star. However, that probe is not necessarily the closest to the final star, yet as long as it reaches its destination star first, the second probe will not consider a visit to the last star. In this case we have artificially extended the exploration time of the probes by prohibiting the second probe from going onto the last star, even though it is much closer and would get there in less time.  This is an instance of a \emph{race condition} or \emph{race hazard}, which occurs in electronic systems where incorrect sequencing of events can result in unwanted delays \citep{Karam1990}.

This underlines a logistical issue in stellar exploration.  We have allowed probes a great deal of autonomy in their exploration; the fleet is not broken up into groups with well-defined areas of exploration, as is the case with the non-replicating studies of \citet{Bjork2007} and \citet{Cotta2009}.  There is no chain of command in the fleet, the fleet continues to grow, and the  probes will act according to simple programming.  Also, the assumption of perfect information is not well motivated.  For this assumption to be valid, the travel time for communication between probes must be significantly lower than the travel time for probes to move towards each other.  In the simulation, the travel time for communication is zero; in reality, communication between the fleet will proceed at the speed of light, $c$.  As the probes achieve speeds of around 0.01 $c$ in the course of exploring the box, and the stellar density is sufficiently high, perfect information is not strictly possible.

\subsection{Other Limitations of the Model}

\noindent We share some of the limitations of the model described in \citet{probe_staticbox}, e.g. not evolving the positions of the stars despite their possessing non-zero velocity vectors.  As velocity vectors will change during the motion of the stars in a gravitational potential, the ability of the maxspeed algorithm to select a star that will have a favourable velocity vector when a probe arrives at it depends significantly on the travel time - stars are much less likely to maintain a similar velocity vector for increasing lengths of time.  We again ignore relativistic effects in this analysis, as they are not significant - for speeds of 0.01$c$, the Lorentz factor $\gamma=1.00005$, showing that classical physics is an acceptable approximation.  This limiting speed of 0.01$c$ is a consequence of the maximum $\Delta v$ being linked to the maximum value of the rotation angle $\delta$ that the velocity vector moves through during the slingshot:

\begin{equation}
\Delta v_{max} = \frac{u^2_{esc}}{\frac{u^2_{esc}}{2u_i} + u_i},
\end{equation}

\noindent Where 

\begin{equation}
u_{esc} = \sqrt{\frac{2GM_*}{R_*}}
\end{equation}

\noindent and we assume solar values for the star mass $M_*$ and radius $R_*$.  Compact objects such as neutron stars and black holes may provide slightly better $\Delta v_{max}$ \citep{Dyson1963}, although it is unclear how such a craft may deal with the extremely strong tidal forces upon its hull.  Besides, any analysis of these types of slingshots must of course consider general relativistic effects, which we do not.

\subsection{Are Advanced Orbital Dynamics Necessary?}

\noindent We have seen that the act of adding self-replication to a probe reduces the difference in total travel time between the powered and slingshot cases significantly (Figure \ref{fig:single_replicate_compare}).  It is therefore important to ask, ``is it worthwhile bothering to program probes to carry out complicated orbital maneouvres at all?''

We would argue the following points: firstly, while there is now only a factor of a few difference in travel time, this is still significant if the exploration is expanded to the entire Galaxy.  We have limited our study to 200,000 stars - increasing that number by a factor of a million (i.e. $10^{11}$ stars) would mean that a factor of a few becomes much more important.  Also, if we consider the trend of total exploration time with star number (Figures \ref{fig:powered_percent} and \ref{fig:slingshot_percent}), then (if we take 99\% of the stars to be an acceptable goal), we can see that the slingshot algorithm scales much better with star number than the powered algorithm, and it seems the difference in travel time will continue to increase as the total number of stars increases.  Assuming powerlaws for how both cases scale with star number, a rough extrapolation suggests that a fleet using the powered algorithm would take around $10^8$ years to explore the Galaxy, compared to around $10^{7}$ for a fleet using the slingshot algorithm.

Even if this is not the case, and there is not a strong difference in travel time on Galactic scales, there is a strong economic argument for the slingshot algorithm.  The massive savings in fuel across the entire fleet by avoiding using powered flight cannot be underestimated (although it is possible that alternative renewable fuel sources such as solar energy may be used).  Being able to achieve a higher velocity may help the process of collecting raw materials from the interstellar medium during transit to the next star.  While the total amount of material collected depends only on the column density of material and path length, the ability to compress the material more efficiently, to conserve space, would benefit from the higher velocities that slingshot trajectories can offer.

\subsection{Improvements to the Model}

Future work should include resolution of the race condition problem described above. This could be achieved in a number of ways, the simplest of which could be to include a travel time limit on the probes. If the nearest unvisited star to them would take longer to travel to than this preprogrammed limit, the probe will simple cease its exploration and will stay on its current star and not produce a replica. This could have the effect of reducing the overall travel time, by preventing probes choosing a destination star at a great distance to themselves, instead of leaving the stars for other probes that are nearby. Another possible way would be for the probes to monitor the position of all other probes and to implement a priority system where the probe closest to the unvisited star is assigned.

A related improvement to these simulations would be to make the probes' information transfer realistic.  When a star becomes marked as visited, there would be a delay in this information reaching the probes, proportional to the distance of the star from members of the probe fleet. This could be simulated by assuming each probe leaves a beacon that emits a signal showing it has been visited, where the signal would propagate throughout space at the speed of light. Probes would then update their own local catalogues of visited stars when they come within the range of the signal.  Alternatively, probes could leave behind their travel history on all visited stars in some form of data box. This box would also contain information on where the probe is next travelling to. Any probe that then later visits an already visited star would sync with the data box, with both entities merging their catalogue data as to what stars have been visited. This way the probes exchange information when they encounter already visited stars. This could help reduce travel time, as probes will avoid stars that have already been visited by other probes with which they have crossed paths.

\section{Conclusions \label{sec:conclusions}}

We have presented results from Monte Carlo Realisation (MCR) simulations of a fleet of self-replicating probes exploring a region of the Galaxy, investigating the effect of slingshot orbital dynamics on the total exploration timescale.  These are the first numerical simulations that explicitly model self-replicating probes undergoing realistic orbital trajectories in a stellar population reminiscent of the Milky Way.  We build on the single probe results of \citet{probe_staticbox}, and we explore the same three scenarios for probe flight: moving to the nearest neighbour under powered flight, moving to the nearest neighbour using slingshot trajectories, and moving to the star that provides the greatest velocity boost via slingshots.  

We find that nearest-neighbour, slingshot-assisted flight is the most time effective exploration method for self-replicating probes (as is the case with a single non-replicating probe). Due to our assumption of perfect information exchange amongst the fleet, the last 1\% of stars take much longer to explore due to ``race conditions'' being set up within the probe network, which require more sophisticated decision making algorithms to avoid.  Despite this, our results confirm that a fleet of self-replicating probes can explore the Galaxy in a timescale commensurate with those normally assumed when posing the Fermi Paradox \citep{Hart1975}, with powered flight at the upper limits of the timescale and slingshot flights at the lower end.  Both are still orders of magnitude less than the age of the Earth, proving that the question underlying the Fermi Paradox is well-posed.

\bibliographystyle{mn2e} 
\bibliography{IJAanicholson}
  
  \end{document}